\documentclass[final,5p,times,twocolumn]{elsarticle}

\usepackage[T1]{fontenc}
\usepackage[utf8]{inputenc}

\usepackage{amsmath}
\usepackage{amsfonts}
\usepackage{amssymb}
\usepackage{amsxtra}
\usepackage{array}
\usepackage{color}
\usepackage{dcolumn}
\usepackage{enumitem}
\usepackage{graphicx}
\usepackage{hepunits}
\usepackage{xspace}
\usepackage{CJKutf8}

\definecolor{purple}{rgb}{0.5,0,0.5}
\definecolor{blue}{rgb}{0.0,0,0.9}
\definecolor{prdblue}{rgb}{0.133,0.118,0.498}
\usepackage[colorlinks=true, pdfstartview=FitV, linkcolor=prdblue, citecolor= prdblue, urlcolor=prdblue]{hyperref}

\usepackage[mathscr,scaled=1.15]{urwchancal}
\DeclareFontFamily{OT1}{pzc}{}
\DeclareFontShape{OT1}{pzc}{m}{it}%
{<-> s * [1.15] pzcmi7t}{}
\DeclareMathAlphabet{\mathpzc}{OT1}{pzc}{m}{it}

\biboptions{sort&compress}

\journal{Physics Letters B}

\begin{document}
\begin{CJK}{UTF8}{song}

\begin{frontmatter}

\title{$\,$\\[-7ex]\hspace*{\fill}{\normalsize{\sf\emph{Preprint no}. NJU-INP 077/23}}\\[1ex]
Constraining the pion distribution amplitude using Drell-Yan reactions on a proton}

\author[NJU,INP]{H.-Y.~Xing
    $\,^{\href{https://orcid.org/0000-0002-0719-7526}{\textcolor[rgb]{0.00,1.00,0.00}{\sf ID}}}$}

\author[HZDR]{M.~Ding
    $^{\href{https://orcid.org/0000-0002-3690-1690}{\textcolor[rgb]{0.00,1.00,0.00}{\sf ID}},}$}

\author[NJU,INP]{Z.-F.~Cui
    $\,^ {\href{https://orcid.org/0000-0003-3890-0242}{\textcolor[rgb]{0.00,1.00,0.00}{\sf ID}}}$}

\author[INP]{A.\,V.~Pimikov%
       $^{\href{https://orcid.org/0000-0002-2874-0196}{\textcolor[rgb]{0.00,1.00,0.00}{\sf ID}},}$}

\author[NJU,INP]{C.\,D.~Roberts%
       $^{\href{https://orcid.org/0000-0002-2937-1361}{\textcolor[rgb]{0.00,1.00,0.00}{\sf ID}},}$}

\author[HZDR,RWTH]{S.\,M.~Schmidt%
$^{\href{https://orcid.org/0000-0002-8947-1532}{\textcolor[rgb]{0.00,1.00,0.00}{\sf ID}},}$}

\address[NJU]{
School of Physics, Nanjing University, Nanjing, Jiangsu 210093, China}
\address[INP]{
Institute for Nonperturbative Physics, Nanjing University, Nanjing, Jiangsu 210093, China}
\address[HZDR]{
Helmholtz-Zentrum Dresden-Rossendorf, Bautzner Landstra{\ss}e 400, D-01328 Dresden, Germany}
\address[RWTH]{
RWTH Aachen University, III. Physikalisches Institut B, D-52074 Aachen, Germany
\\[1ex]
%
\href{mailto:m.ding@hzdr.de}{m.ding@hzdr.de} (MD);
\href{mailto:phycui@nju.edu.cn}{phycui@nju.edu.cn} (ZFC);
\href{mailto:cdroberts@nju.edu.cn}{cdroberts@nju.edu.cn} (CDR)
%
\\[1ex]
Date: 2023 September 05 \\[-6ex]
}

\begin{abstract}
Using a reaction model that incorporates pion bound state effects and continuum results for proton parton distributions and the pion distribution amplitude, $\varphi_\pi$, we deliver parameter-free predictions for the $\mu^+$ angular distributions in $\pi N \to \mu^+ \mu^- X$ reactions on both unpolarised and polarised targets.  The analysis indicates that such angular distributions are sensitive to the pointwise form of $\varphi_\pi$ and suggests that unpolarised targets are practically more favourable.  The precision of extant data is insufficient for use in charting $\varphi_\pi$; hence, practical tests of this approach to charting $\varphi_\pi$ must await data with improved precision from new-generation experiments.  The reaction model yields a nonzero single-spin azimuthal asymmetry, without reference to $T$-odd parton distribution functions (DFs).  This may necessitate additional care when attempting to extract such $T$-odd DFs from data.
\end{abstract}


\end{frontmatter}
\end{CJK}

\section{Introduction}
%
A new era is dawning, in which high-luminosity, high-energy accelerators will enable science to escape the limitations of a hundred-year focus on the proton and begin to probe deeply into the structure of the pion \cite{Sawada:2016mao, Adams:2018pwt, Aguilar:2019teb, Chen:2020ijn, Anderle:2021wcy, Arrington:2021biu, Quintans:2022utc, Accardi:2023chb}, Nature's most fundamental Nambu-Goldstone boson \cite{Horn:2016rip, Roberts:2021nhw}.  A basic expression of that structure is the pion's parton distribution amplitude (DA), $\varphi_\pi(x;\zeta)$, which plays a key role in perturbative analyses of hard exclusive processes in quantum chromodynamics (QCD) \cite{Lepage:1979zb, Efremov:1979qk, Lepage:1980fj}.  As the probability amplitude for finding a valence quark carrying light-front fraction $x$ of the pion's total momentum at a resolving scale $\zeta$, $\varphi_\pi(x;\zeta)$ features in an array of leading-order perturbative-QCD hard-scattering formulae for elastic processes that must provide an accurate approximation on some domain $\zeta > \zeta_{\rm hard} > m_p$, where $m_p$ is the proton mass.  The value of $\zeta_{\rm hard}$ is not predicted by perturbative-QCD and can be process dependent.

On $m_p/\zeta \simeq 0$, the DA takes its asymptotic form \cite{Lepage:1979zb, Efremov:1979qk, Lepage:1980fj}:
\begin{equation}
\varphi_{\rm as}(x)=6 x(1-x)\,.
\end{equation}
However, when $\varphi_{\rm as}$ is used in leading-order hard-scattering formulae to estimate cross-sections, the comparison with data is typically poor \cite{Volmer:2000ek, Horn:2006tm, Tadevosyan:2007yd, Blok:2008jy, Huber:2008id, Aubert:2009mc, Uehara:2012ag}.  Such outcomes long ago highlighted a critical question, whose answer lies outside perturbation theory; namely, what is the pointwise form of $\varphi_\pi(x;\zeta)$ at resolving scales relevant to achievable experiments ($\zeta \approx 2\,$GeV$\,=:\zeta_2$)?

Early phenomenology pointed to an answer, \emph{viz}.\ that $\varphi_\pi(x;\zeta_2)$ should be greatly dilated with respect to $\varphi_{\rm as}(x)$ \cite{Chernyak:1983ej}.  However, encumbered by constraints imposed by insisting that $\varphi_\pi(x;\zeta_2)$ be represented as an expansion in eigenfunctions of the DA evolution operator \cite{Lepage:1979zb, Efremov:1979qk, Lepage:1980fj}, with only a few terms, such DAs were typically bimodal or ``double humped'', \emph{i.e}., exhibited two maxima, one each side of a single (deep) minimum at $x=1/2$.

With development of continuum Schwinger function methods (CSMs) \cite{Bashir:2012fs, Eichmann:2016yit, Qin:2020rad} to a point from which $\varphi_\pi$ could directly be calculated \cite{Chang:2013pq}, it became clear that a double-humped form is inconsistent with the character of a ground-state pseudoscalar meson \cite{Li:2016dzv}.  Instead, the DA of the ground-state pion should be a broad concave function.  A similar conclusion has been reached via other means -- see, \emph{e.g}., Refs.\,\cite{Brodsky:2006uqa, Segovia:2013eca, Gao:2016jka, Zhang:2017bzy, Qian:2020utg, Choi:2020xsr, Mondal:2021czk}.

Following the discussion in Ref.\,\cite[Sec.\,3B]{Roberts:2021nhw} and adapting the representation in Ref.\,\cite{Cui:2022bxn}, the prediction in Ref.\,\cite{Chang:2013pq} can be written in the following compact form:
\begin{equation}
\label{pionDA}
\varphi_\pi(x;\zeta_2) = {\mathpzc n}_0 \ln (1 + x (1 - x)/{\mathpzc r}_\varphi^2 )\,,\quad {\mathpzc r}_\varphi= 0.162\,,
\end{equation}
with ${\mathpzc n}_0$ ensuring unit normalisation.  The dilation is now understood as a corollary of emergent hadron mass (EHM) \cite{Roberts:2020hiw, Roberts:2020udq, Roberts:2021xnz, Roberts:2021nhw, Binosi:2022djx, Ding:2022ows, Ferreira:2023fva, Carman:2023zke}, a nonperturbative feature of quantum chromodynamics (QCD).

Commonly considered hard scattering formulae for exclusive processes involving pions are only sensitive to the $1/x$-moment of $\varphi_\pi$ \cite{Lepage:1980fj}.  Consequently, whilst related experiments can contribute to a discussion of its dilation \cite{E12-19-006}, they cannot serve as keen discriminators between contrasting pointwise forms of $\varphi_\pi$.  On the other hand, it has been argued \cite{Brandenburg:1994wf} that the angular distribution of the $\mu^+$-lepton produced in the Drell-Yan process \cite{Drell:1970wh, Holt:2010vj}: $\pi^- + p \to \mu^+ + \mu^- + X$, where $X$ is a shower of undetected hadrons, is sensitive to the shape of $\varphi_\pi(x)$, especially if the proton involved is longitudinally polarised \cite{Brandenburg:1995pk, Bakulev:2007ej}.  We examine these possibilities herein, in the context of contemporary and anticipated data \cite{Anderle:2021wcy, Arrington:2021biu, Quintans:2022utc, Accardi:2023chb}.

\begin{figure*}[t]
\centerline{%
\includegraphics[clip, width=0.9\textwidth]{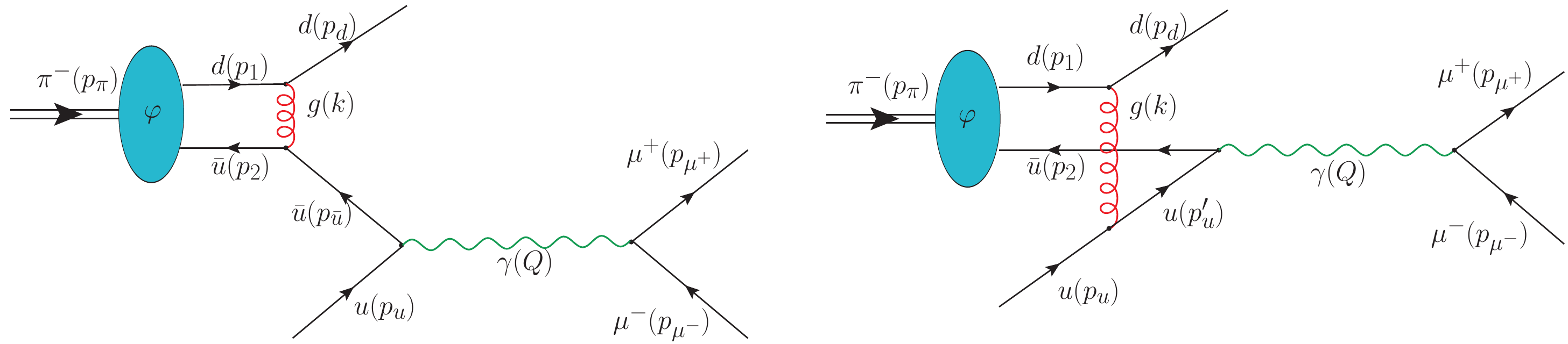}}
\caption{\label{DrellYan}
Contributions supposed to describe the production of $\mu^+ \mu^-$ pairs with large invariant mass, $Q^2$, and large longitudinal-momentum fraction, $x_L$, in the process $\pi^-(p_\pi) N(p_N) \to \mu^+ \mu^- X$.
With $s=(p_\pi+p_N)^2$, $\tau=Q^2/s$, and $Q_{L/T}$ being the components of the (virtual) photon momentum that are parallel/perpendicular to $p_\pi$, then the $\mu^+ \mu^-$ longitudinal momentum fraction is $x_L=2 Q_L/\surd s$.  $x_L$ takes the maximum value $x_L^{\rm max}=1-(Q^2+Q_T^2)/s$, which is typically slightly less than unity on the relevant kinematic domain.
$\pi$ and $N$ longitudinal momentum fractions are $x_{\pi,N} = [\pm x_F +(x_F^2+4 \tau^2)^{1/2}]/2$: $x_F = x_\pi - x_N$; $x_\pi x_N = \tau$.
%
%
}
\end{figure*}

\section{Drell-Yan pair angular distributions}
\label{SecFormulae}
In $\pi^\pm N$ scattering, it is typically supposed that the principal mechanism behind the production of lepton pairs with invariant mass, $Q^2$, and longitudinal momentum fraction, $x_L$, both large, is the Drell-Yan process \cite{Drell:1970wh, Holt:2010vj}.  A treatment of this process that incorporates pion bound state effects is depicted in Fig.\,\ref{DrellYan}.
Supposing that parton transverse momentum is small, \emph{i.e}., $\rho^2 = Q_T^2/Q^2 \simeq 0$, and writing $p_{\bar u} = x_{\bar u} p_\pi$, then $x_{\bar u}\approx x_\pi$, so the process is sensitive to the pion valence antiquark distribution.  Similarly, under these assumptions $p_u = x_u P_N$ and $x_u \approx x_N$.
As drawn in Fig.\,\ref{DrellYan}, owing to the hard gluon exchange between the annihilating valence antiquark in the pion and the spectator valence quark (and the partner process), the associated cross-section is sensitive to the pion leading-twist two-particle distribution amplitude, $\varphi_\pi$.  The hard gluon exchange also introduces a dependence on $Q_T$.

Before proceeding, it is worth reiterating some caveats on this description of the Drell-Yan reaction \cite{Brandenburg:1994wf}.
The expression of pion bound state effects in Fig.\,\ref{DrellYan} involves a mixture of nonperturbative and perturbative phenomena.
Given the kinematics of the problem, it should be reasonable on the valence antiquark domain; and, in this case, one may anticipate that additional hard gluon exchanges are suppressed.
Soft gluon contributions are assumed to be expressed in evolution of the pion DA.
Notably, nucleon bound state effects are not incorporated in the same way as those relating to the pion.  One justifies this by recalling the kinematic focus on low $x_u$.
Gluon emissions to the final state are assumed to be incorporated in evolution of the proton DFs with hard emissions suppressed by powers of the strong running coupling.
No systematic tests of these assumptions are available.

Using angles defined in the Gottfried-Jackson frame, with the $\pi^-$ beam defining the $z$-direction \cite[Fig.\,3]{Bakulev:2007ej}, the five-fold differential cross-section associated with Fig.\,\ref{DrellYan} takes the following form:
{\allowdisplaybreaks
\begin{align}
& \frac{d^5 \sigma(\pi^- N \to \mu^+ \mu^- X)}{dQ^2 dQ_{T}^2 dx_L d\cos\theta d\phi} \propto N(\tilde x, \rho)  \nonumber \\
& \quad \times  \left[1+\lambda \cos^2 \theta + \mu \sin 2\theta \cos\phi + \tfrac{1}{2}\nu \sin^2\theta \cos 2\phi \right. \nonumber \\
&
\qquad \left. + \bar\mu \sin 2\theta \sin\phi + \tfrac{1}{2} \bar\nu \sin^2\theta \sin 2\phi \right]\,,
\label{DYcross}
\end{align}
where the last line contributes when the nucleon target is longitudinally polarised and is otherwise absent, and the masses of the pion and current-quarks are neglected.
}

The angular distribution coefficients are functions of the following kinematic variables:
$Q^2$,
$s=(p_\pi+p_N)^2$,
$\rho^2 = Q_T^2/Q^2$,
{\allowdisplaybreaks
\begin{equation}
\tilde x = x_{\bar u}/(1+\rho^2) = [x_L + (x_L^2+4 \tau [1+\rho^2] )^{1/2}]/[2(1+\rho^2)]\,.
\end{equation}
Namely \cite{Brandenburg:1994wf, Brandenburg:1995pk, Bakulev:2007ej}:
\begin{subequations}
\label{coefficients}
\begin{align}
N & = 2 \left\{ (1-\tilde x)^2
 ( [F_\varphi + {\rm Re} I_\varphi(\tilde x)]^2+[{\rm Im} I_\varphi(\tilde x)]^2) \right. \nonumber \\
 & \qquad \left.+ [4+\rho^2]\rho^2 \tilde x^2 F_\varphi^2 \right\}\,,\\
 & = 2 \left\{ (1-\tilde x)^2 [F_\varphi + {\rm Re} I_\varphi(\tilde x)]^2
 + \pi^2 \varphi(\tilde x;Q^2)^2 \right. \nonumber \\
 & \qquad \left. + [4+\rho^2]\rho^2 \tilde x^2 F_\varphi^2\right\} \,, \label{corrected}\\
 N \lambda & = 2 \left\{
 (1-\tilde x)^2  ( [{\rm Im} I_\varphi(\tilde x)]^2 + [F_\varphi + {\rm Re} I_\varphi(\tilde x)]^2)\right. \nonumber \\
 & \qquad \left. - (4-\rho^2)\rho^2 \tilde x^2 F_\varphi^2 \right\}\,,\\
N \mu & = - 4\rho F_\varphi \tilde x \left\{
(1-\tilde x)[F_\varphi + {\rm Re} I_\varphi(\tilde x)] + \rho^2 \tilde x F_\varphi
\right\}\,,\\
N\nu & = - 8 \rho^2 \tilde x (1-\tilde x) F_\varphi [F_\varphi + {\rm Re} I_\varphi(\tilde x)]\,,\\
N \bar\mu & = - \tilde\mu_N  4\pi {\mathpzc s}_L \, \rho \tilde x F_\varphi \, \varphi(\tilde x; Q^2)
\label{corrected1}\\
\bar\nu & = 2 \rho \bar\mu\,, \label{barnu}\\
\bar\mu_N & = \frac{\tfrac{4}{9} \Delta {\mathpzc u}_V(x_u;\zeta)+\tfrac{4}{9} \Delta {\mathpzc u}_S(x_u;\zeta)+\tfrac{1}{9} \Delta {\mathpzc d}_S(x_u;\zeta)}
{\tfrac{4}{9} {\mathpzc u}_V(x_u;\zeta)+\tfrac{4}{9} {\mathpzc u}_S(x_u;\zeta)+\tfrac{1}{9} {\mathpzc d}_S(x_u;\zeta)}\,, \label{ProtonDFratio}
\end{align}
\end{subequations}
where
\begin{equation}
F_\varphi = \int_0^1 dy \frac{1}{y} \varphi(y, Q^2)\,,\;
I_\varphi(\tilde x) = \int_0^1 dy  \frac{\varphi(y, Q^2)}{y [y+\tilde x -1 + i \epsilon]}\,,
\end{equation}
${\mathpzc s}_L = \pm 1$ expresses the target's longitudinal polarisation state,
and
$\Delta {\mathpzc q}_{V/S}(x;\zeta)$, ${\mathpzc q}_{V/S}(x;\zeta)$ are, respectively, light quark polarised and unpolarised valence/sea quark parton distribution functions (DFs) within the proton at resolving scale $\zeta$.  Note that the reaction is only sensitive to these DFs when the target is polarised.
}

\begin{figure*}[t]
\hspace*{-1ex}\begin{tabular}{lll}
{\sf A} & {\sf B}  & {\sf C}  \hspace*{\fill} $\rho=0.2$\\[-0ex]
\includegraphics[clip, width=0.315\textwidth]{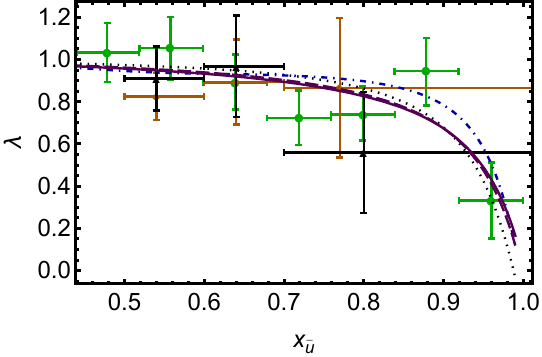} &
\includegraphics[clip, width=0.325\textwidth]{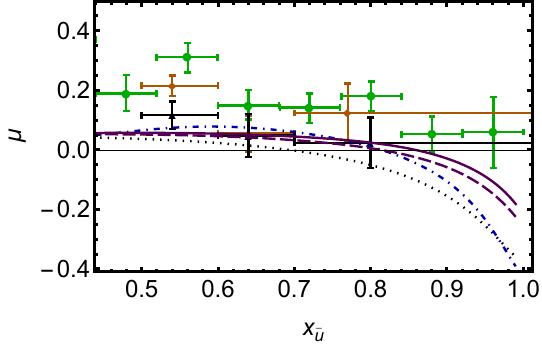} &
\includegraphics[clip, width=0.315\textwidth]{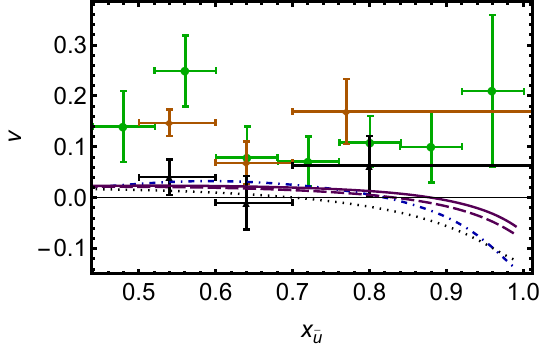} \\
{\sf D} & {\sf E}  & {\sf F} \hspace*{\fill} $\rho=0.4$  \\[-0ex]
\includegraphics[clip, width=0.315\textwidth]{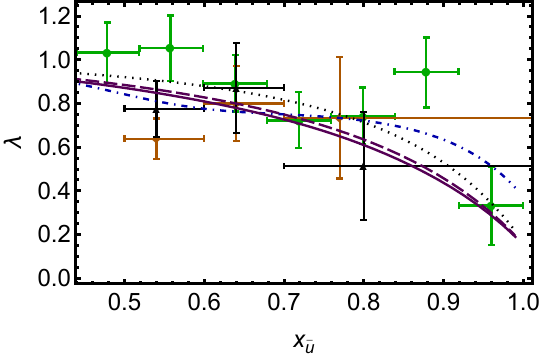} &
\includegraphics[clip, width=0.325\textwidth]{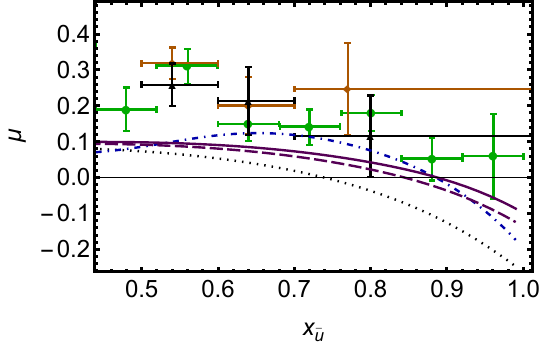} &
\includegraphics[clip, width=0.315\textwidth]{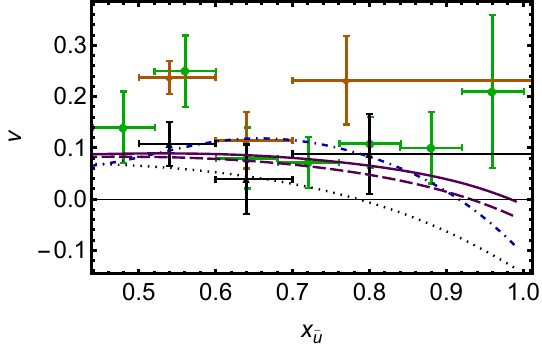}
\end{tabular}
\caption{\label{ResultsUP02}
Results for the unpolarised-target angular distribution coefficients in Eq.\,\eqref{DYcross}.
Row~1 (panels {\sf A}, {\sf B}, {\sf C}) -- $\rho=0.2$;
Row~2 (panels {\sf D}, {\sf E}, {\sf F}) -- $\rho=0.4$.
The predictions are insensitive to the proton DFs, which cancel via normalisation.
%
\emph{Legend}.
Solid purple curve: our predictions with CSM pion DA evolved from $\zeta_2=2\,$GeV to $\zeta_5$.
Long-dashed purple: our predictions with pion DA evolved from $\zeta_1=1\,$GeV.
Dot-dashed blue: results obtained using the pion DA discussed in Ref.\,\cite[BMS]{Bakulev:2001pa}, which is associated therein with initial scale $\zeta_1$.
Dotted black: results obtained with $\varphi_{\rm as}(x)$.
Results inferred from data, all associated with
$Q^2 = \zeta_{5}^2$:
orange diamonds -- Ref.\,\cite[NA10]{Guanziroli:1987rp} -- 194\,GeV $\pi^-$ beams;
black triangles -- Ref.\,\cite[NA10]{Guanziroli:1987rp} -- 286\,GeV $\pi^-$;
green circles -- Ref.\,\cite[E615]{Conway:1989fs} -- 252\,GeV $\pi^-$.
%
%
}
\end{figure*}

Equations~\eqref{corrected}, \eqref{corrected1}, \eqref{barnu} correct an error in Ref.\,\cite[Eq.\,(2.12)]{Brandenburg:1995pk}, \emph{viz}.\ in the denominator there, one finds $(1-\tilde x)^2 \pi^2 \varphi(\tilde x; Q^2)^2$ in place of the correct result, $\pi^2 \varphi(\tilde x; Q^2)^2$.  Notwithstanding this, we have established that the correct formula was used in Ref.\,\cite{Brandenburg:1995pk} to produce the numerical results.  On the other hand, Ref.\,\cite{Bakulev:2007ej} used the incorrect form in calculations relating to longitudinally polarised protons.  Hence, the numerical results drawn in Ref.\,\cite[Figs.\,6, 7, 9, 10]{Bakulev:2007ej} are incorrect.  Further, the curves in Ref.\,\cite[Fig.\,8]{Bakulev:2007ej} have the wrong sign.

\section{Comparisons with data -- unpolarised target}
\label{SecUP}
Three sets of pion + nucleus Drell-Yan data are available:
Ref.\,\cite[NA10]{Guanziroli:1987rp} -- 194\,GeV and 286\,GeV $\pi^-$ beams on $W$ target; and
Ref.\,\cite[E615]{Conway:1989fs} -- 252\,GeV $\pi^-$ on $W$.
The NA10 data are reported in the Collins-Soper frame.  They can be mapped into the Gottfried-Jackson frame using formulae in Ref.\,\cite[Appendix E]{Conway:1989fs}.  The mapping is $\rho^2$-dependent.
All experiments may be identified with mean invariant mass squared $Q^2 = (5.2\,{\rm GeV})^2=:\zeta_{5}^2$.

Using the formulae in Sec.\,\ref{SecFormulae}, comparison with such data becomes possible once the proton DFs and pion DA are known.  Previous such comparisons have used DFs fitted to other data and models for the pion DA.  Herein, profiting from recent progress made using continuum Schwinger function methods (CSMs), which has delivered unified predictions for all relevant DFs \cite{Chang:2022jri, Lu:2022cjx, Cheng:2023kmt} and the pion DA \cite{Chang:2013pq, Cui:2020tdf}, we provide parameter-free predictions for comparison with Drell-Yan data.  Consequently, such data becomes a test of the EHM paradigm established by modern studies of strong QCD \cite{Roberts:2021nhw, Binosi:2022djx, Ding:2022ows, Ferreira:2023fva, Carman:2023zke}.

\begin{figure}[t]
\centerline{\includegraphics[clip, width=0.42\textwidth]{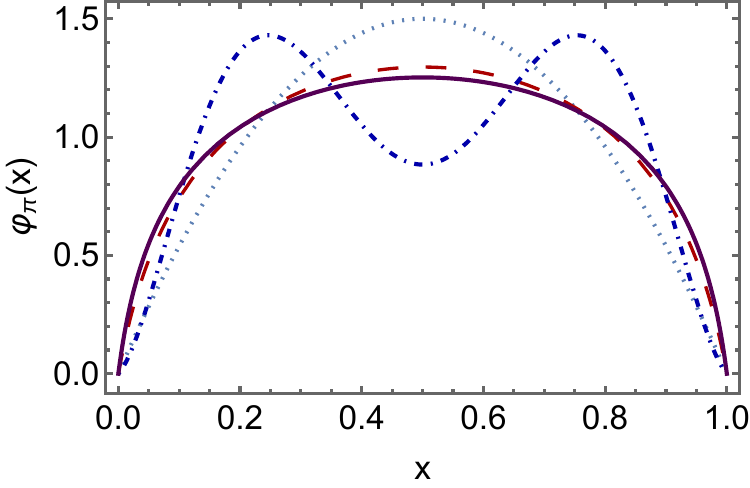}}
\caption{\label{DAcompare}
Pion DAs used herein.
Solid purple curve -- CSM prediction \cite{Chang:2013pq}, \emph{viz}.\ the concave and dilated function in Eq.\,\eqref{pionDA};
dot-dashed curve -- Ref.\,\cite[BMS]{Bakulev:2001pa}, evolved to $\zeta_2$;
and dotted black curve -- $\varphi_{\rm as}(x)$.
As explained elsewhere \cite[Sec.\,3B]{Roberts:2021nhw}, the information used in Ref.\,\cite{Bakulev:2001pa} to obtain the BMS DA could equally have been employed to recover a pion DA in the form of Eq.\,\eqref{pionDA}, in which case one finds ${\mathpzc r}_\varphi^{\rm BMS}=0.226$ and a 4\% root-mean-square difference between the $(1-2x)^{2,4}$ Mellin moments of the original and revised curves.  This difference is within the original BMS uncertainty.  The revised DA is drawn as the long-dashed red curve.  It is practically indistinguishable from the CSM prediction.
}
\end{figure}

Our results for the unpolarised-target angular distribution coefficients in Eq.\,\eqref{DYcross} are exemplified in Fig.\,\ref{ResultsUP02}.
The solid purple curve in each panel was obtained using
the pion DA in Ref.\,\cite{Chang:2013pq}, \emph{viz}.\ the concave and dilated function expressed in Eq.\,\eqref{pionDA}, drawn as the solid purple curve in Fig.\,\ref{DAcompare}.
As noted in Eq.\,\eqref{pionDA}, this DA is reported at $\zeta_2$, wherefrom we evolved it to the experiment scale, $\zeta_5$, by adapting the all-orders DGLAP evolution scheme detailed in Ref.\,\cite{Yin:2023dbw} to ERBL evolution \cite{Lepage:1979zb, Efremov:1979qk} and using $\zeta_2$ as the initial scale.

The rows in Fig.\,\ref{ResultsUP02} compare two values of $Q_T^2$, expressed via $\rho = 0.2, 0.4$, which are the mean values of this kinematic variable for NA10 and E615 data, respectively.
Evidently, within uncertainties, available data are compatible with our predictions.  On the other hand, as highlighted by the dot-dashed blue and dotted black curves in Fig.\,\ref{ResultsUP02}, results inferred from existing NA10 data are of insufficient precision to discriminate between different forms of the pion DA.

On the other hand, there is significance in comparisons with the E615 data in Fig.\,\ref{ResultsUP02}.
Using the CSM DA \cite{Chang:2013pq}, one calculates  $\chi^2 / {\rm datum} = 2.7$ over all directly relevant panels in Fig.\,\ref{ResultsUP02}, \emph{viz}.\ panels {\sf D}\,--\,{\sf F}.
In contrast, using the bimodal BMS DA \cite{Bakulev:2001pa}, one finds $\chi^2 / {\rm datum} = 2.3$;
and $\varphi_{\rm as}(x)$ yields $\chi^2 / {\rm datum} = 4.8$.

The pointwise forms of these three DAs are compared in Fig.\,\ref{DAcompare}.  Contemporary simulations of lattice-regularised QCD produce results that favour the CSM DA profile -- see, \emph{e.g}., the discussion in Ref.\,\cite[Secs.\,3B, 8D]{Roberts:2021nhw}.
The BMS DA is bimodal: developed subject to the constraint that $\varphi_\pi(x;\zeta_2)$ be represented via an expansion in eigenfunctions of the DA evolution operator \cite{Lepage:1979zb, Efremov:1979qk, Lepage:1980fj}, the dilation -- $x\simeq 0,1$ endpoint enhancement \emph{cf}.\ $\varphi_{\rm as}(x)$ -- required for agreement with observables comes at the cost of a deep local minimum at $x=1/2$.
%
%
Given the distinct $x$-dependences of these DAs, the comparisons drawn herein provide a robust illustration of the sensitivity of $\pi N$ Drell-Yan processes to the pion DA when the reaction is assumed to proceed as drawn in Fig.\,\ref{DrellYan}.

Considering each of the panels in the second row of Fig.\,\ref{ResultsUP02} independently, one finds, using the CSM, BMS and asymptotic DAs respectively,
%
{\sf D} -- $\chi^2 / {\rm datum} = 1.9$, $1.4$, $1.0$;
{\sf E} -- $\chi^2 / {\rm datum} = 4.8$, $3.9$, $10$;
{\sf F} -- $\chi^2 / {\rm datum} = 1.3$, $1.5$, $3.5$.
Plainly, at modest to moderate $\rho$, $\mu$ and $\nu$ provide greatest sensitivity to the pion DA because of their explicit $\rho$ dependence and potential amplifications owing to DA dilation -- see Fig.\,\ref{DAcompare}.
Moreover, it is clear from Fig.\,\ref{ResultsUP02} that extant data prefer a dilated pion DA.  Nevertheless, significantly improved precision in data on $\pi N$ Drell-Yan from unpolarised targets is required before they could be used to discriminate between the pointwise behaviour of realistic pion DAs, \emph{i.e}., inequivalent functions that are consistent, \emph{e.g}., with data on the $\gamma^\ast \gamma \to \pi^0$ transition form factor \cite{Stefanis:2012yw, Raya:2015gva}.

\emph{A priori}, the scale to be associated with the BMS DA is unknown.  It is typically chosen to be $\zeta_1$.  Thus in order to identify core differences between cross-sections produced by the BMS and CSM DAs, Fig.\,\ref{ResultsUP02} also depicts results obtained with the CSM DA evolved $\zeta_1 \to \zeta_5$.  Plainly, the starting scale is not the source of differences between results based on the BMS and CSM DAs.

\begin{figure}[t]
%
\centerline{\includegraphics[clip, width=0.42\textwidth]{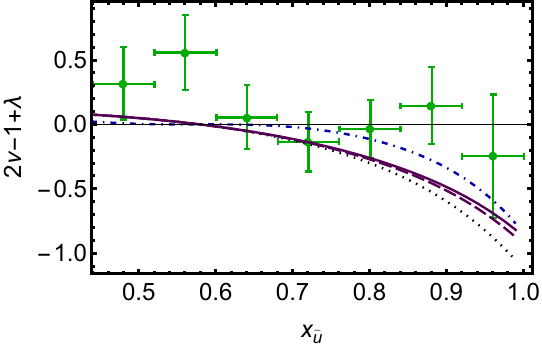}}
\caption{\label{FLamTung}
Lam-Tung combination of angular distribution coefficients.
\emph{Legend}.
Solid purple curve: our predictions with pion DA evolved $\zeta_2 \to \zeta_5$.
Long-dashed purple curve: our predictions with pion DA evolved $\zeta_1 \to \zeta_5$.
Dot-dashed blue curve: results obtained using the pion DA discussed in Ref.\,\cite[BMS]{Bakulev:2001pa}, associated therein with initial scale $\zeta_1$.
Dotted black curve: results obtained with $\varphi_{\rm as}(x)$.
Results inferred from data, associated with a mean invariant mass squared $Q^2 = \zeta_{5}^2$:
green circles -- Ref.\,\cite[E615]{Conway:1989fs}.
}
\end{figure}

Figure~\ref{FLamTung} displays the $x_{\bar u}$-dependence of the Lam-Tung combination of angular distribution coefficients \cite{Lam:1980uc}: $2 \nu - 1 + \lambda$, determined from E615 data at the average value of $\rho=0.4$.  Analogous to the Callan-Gross relation in deep inelastic scattering, this combination is zero in a parton model treatment of the Drell-Yan process \cite{Holt:2010vj}.  Based on this image alone, one cannot conclude that the Lam-Tung relation is violated: the data are consistent with zero at the level of 55\%.  On the other hand, violation is apparent in the other two panels of Ref.\,\cite[Fig.\,38]{Conway:1989fs}.  The reaction model considered herein produces a violation of the Lam-Tung relation ($2 \nu - 1 + \lambda=0$) because it incorporates parton transverse momentum via $\rho \neq 0$.  Nonetheless, distinctions between predictions obtained using different pion DAs are small in comparison with the precision of extant data.

\begin{figure*}[t]
\hspace*{-1ex}\begin{tabular}{lll}
{\sf A} & {\sf B}  & {\sf C}  \hspace*{\fill} $\rho=0.2$\\[-0ex]
\includegraphics[clip, width=0.325\textwidth]{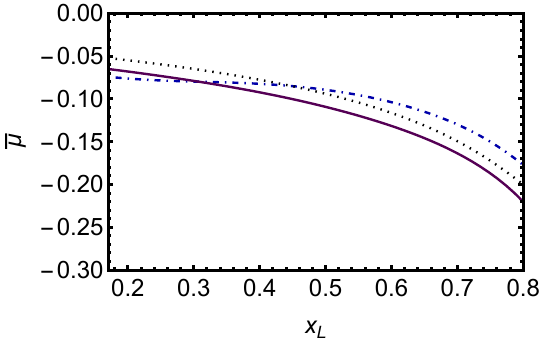} &
\includegraphics[clip, width=0.325\textwidth]{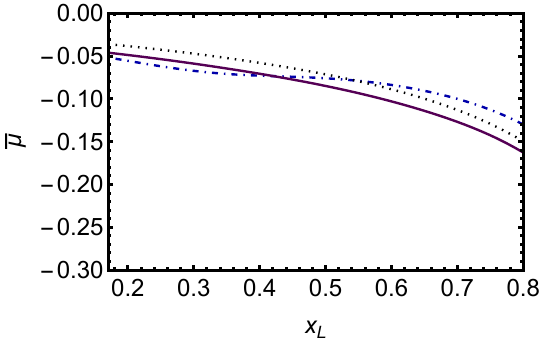} &
\includegraphics[clip, width=0.305\textwidth]{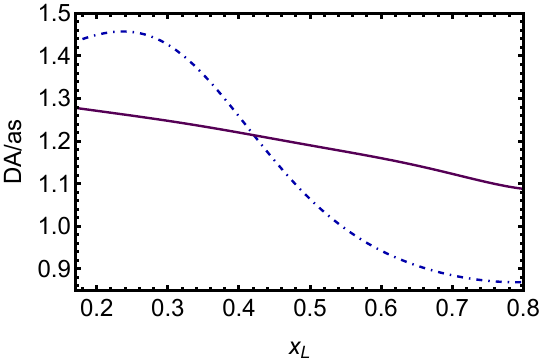} \\
{\sf D} & {\sf E}  & {\sf F} \hspace*{\fill} $\rho=0.4$  \\[-0ex]
\includegraphics[clip, width=0.325\textwidth]{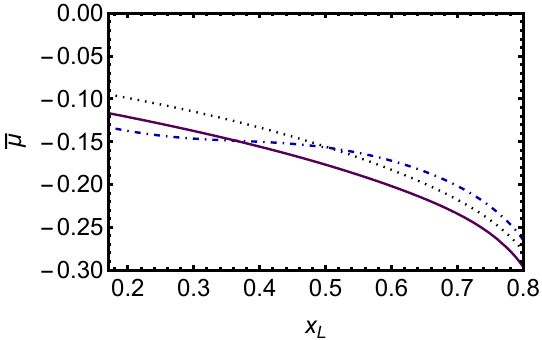} &
\includegraphics[clip, width=0.325\textwidth]{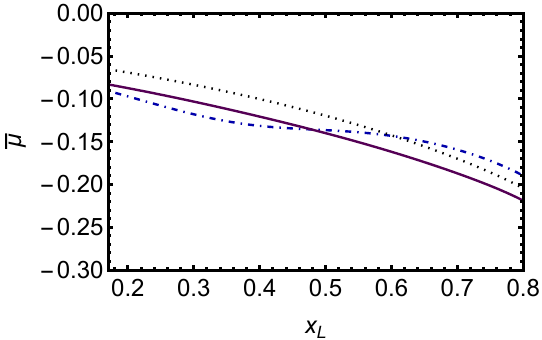} &
\includegraphics[clip, width=0.305\textwidth]{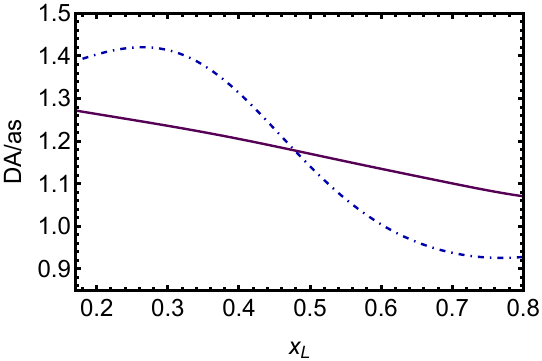}
\end{tabular}
\caption{\label{ResultsPol}
Results for the polarised-target angular distribution coefficient $\bar\mu$ in Eq.\,\eqref{DYcross}.
The predictions are sensitive to the helicity-dependent/helicity-independent proton DF ratio -- see Eq.\,\eqref{ProtonDFratio}.
Recall: $x_L = 2 Q_L/\surd s = x_{\bar u}- [1+\rho^2] \tau/x_{\bar u}$.
Row~1 (panels {\sf A}, {\sf B}, {\sf C}) -- $\rho=0.2$;
Row~2 (panels {\sf D}, {\sf E}, {\sf F}) -- $\rho=0.4$.
Column~1 (panels {\sf A}, {\sf D}) -- $s=200\,{\rm GeV}^2$;
Column~2 (panels {\sf B}, {\sf E}) -- $s=357\,{\rm GeV}^2$.
Column~3 (panels {\sf C}, {\sf F}) -- $s=357\,{\rm GeV}^2$ ratio of results obtained using CSM and asymptotic DA or BMS and asymptotic DA.
%
\emph{Legend}.
Solid purple curve: predictions with CSM DA \cite{Chang:2013pq}, evolved $\zeta_2 \to \zeta_5$.
Dot-dashed blue curve: results obtained using BMS DA \cite{Bakulev:2001pa}, evolved $\zeta_1 \to \zeta_5$.
Dotted black curve: results obtained with $\varphi_{\rm as}(x)$.
\emph{N.B}. $s=357\,{\rm GeV}^2$ is typical of measurements performed by the COMPASS Collaboration and anticipated by the AMBER Collaboration \cite{Quaresma:2017ear, Lien:2019yif, Quintans:2022utc}.
%
%
%
}
\end{figure*}

\begin{figure*}[t]
\hspace*{-1ex}\begin{tabular}{lll}
{\sf A} & {\sf B}  & {\sf C}  \hspace*{\fill} $\rho=0.2$\\[-0ex]
\includegraphics[clip, width=0.32\textwidth]{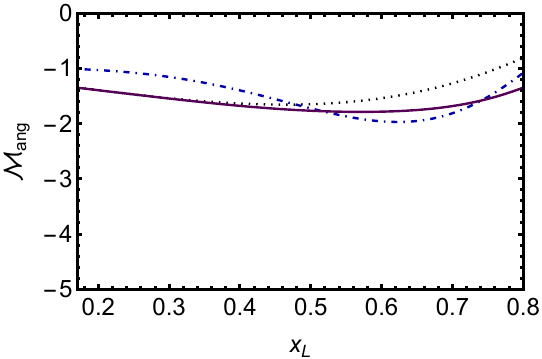} &
\includegraphics[clip, width=0.32\textwidth]{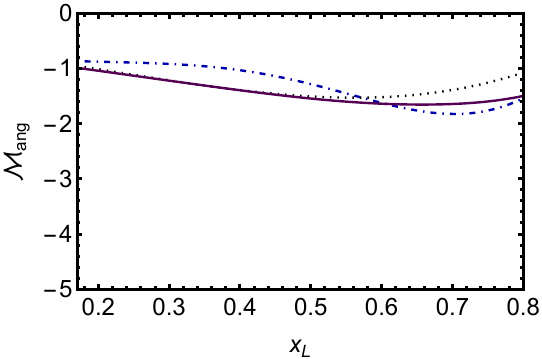} &
\includegraphics[clip, width=0.31\textwidth]{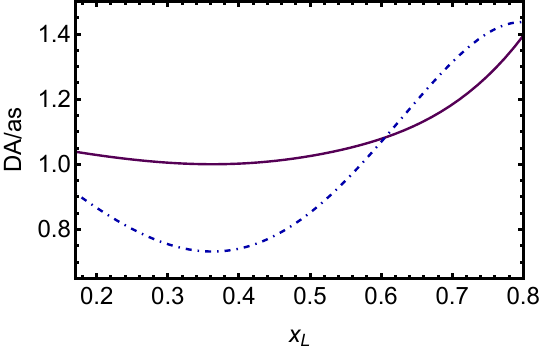} \\
{\sf D} & {\sf E}  & {\sf F} \hspace*{\fill} $\rho=0.4$  \\[-0ex]
\includegraphics[clip, width=0.32\textwidth]{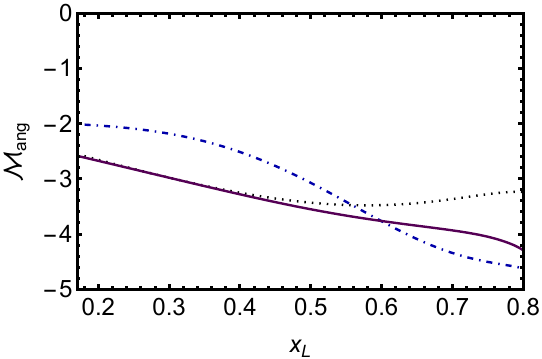} &
\includegraphics[clip, width=0.32\textwidth]{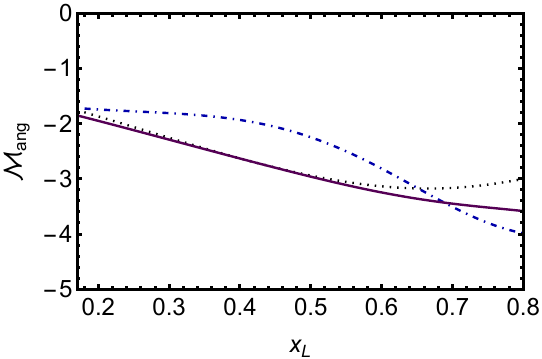} &
\includegraphics[clip, width=0.31\textwidth]{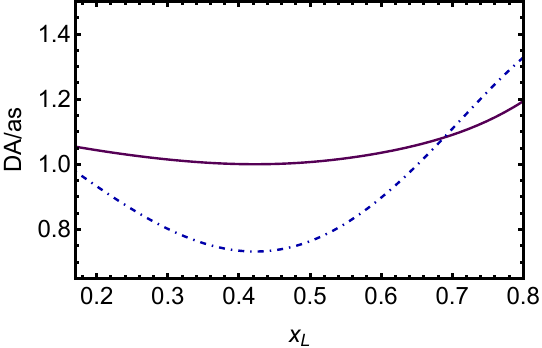}
\end{tabular}
\caption{\label{ResultsMang}
Results for the angular moment, Eq.\,\eqref{EqMang}, of the polarised-target Drell-Yan cross-section.
The predictions are sensitive to the helicity-dependent/helicity-independent proton DF ratio -- see Eq.\,\eqref{ProtonDFratio}.
%
%
Row~1 (panels {\sf A}, {\sf B}, {\sf C}) -- $\rho=0.2$;
Row~2 (panels {\sf D}, {\sf E}, {\sf F}) -- $\rho=0.4$.
Column~1 (panels {\sf A}, {\sf D}) -- $s=200\,{\rm GeV}^2$;
Column~2 (panels {\sf B}, {\sf E}) -- $s=357\,{\rm GeV}^2$.
Column~3 (panels {\sf C}, {\sf F}) -- $s=357\,{\rm GeV}^2$ ratio of results obtained using CSM and asymptotic DA or BMS and asymptotic DA.
%
\emph{Legend}.
Solid purple curve: predictions with CSM pion DA \cite{Chang:2013pq}, evolved $\zeta_2 \to \zeta_5$.
Dot-dashed blue curve: results obtained using BMS DA \cite{Bakulev:2001pa}, evolved $\zeta_1 \to \zeta_5$.
Dotted black curve: results obtained with $\varphi_{\rm as}(x)$.
%
%
%
}
\end{figure*}

\begin{figure}[t]
\vspace*{0ex}

\leftline{\hspace*{0.5em}{{\textsf{A}}}}
\vspace*{-2ex}
\centerline{\includegraphics[width=0.38\textwidth]{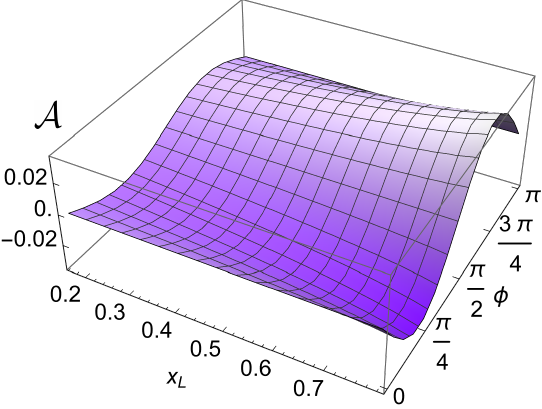}}
\vspace*{2ex}

\leftline{\hspace*{0.5em}{{\textsf{B}}}}
\vspace*{-2ex}
\centerline{\includegraphics[width=0.36\textwidth]{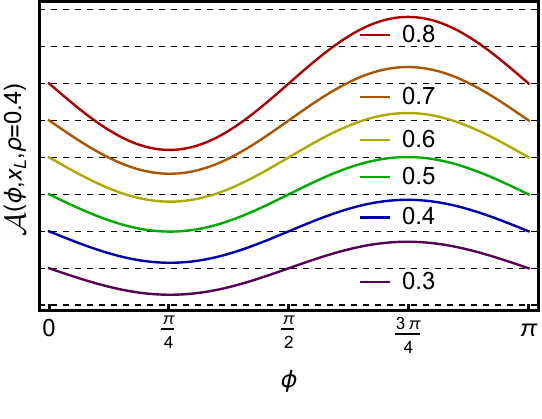}}

\caption{\label{FigSSA}
Our prediction for the SSA in Eq.\,\eqref{ASSA} evaluated using $s=357\,{\rm GeV}^2$, $\rho = 0.4$, the proton DFs in Refs.\,\cite{Lu:2022cjx, Cheng:2023kmt} and the CSM pion DA \cite{Chang:2013pq}, \emph{i.e}., using the inputs that produce the solid purple curve in Fig.\,\ref{ResultsPol}E.
Panel \textsf{A}.  Three-dimensional image.
Panel \textsf{B}.  $\phi$ profiles at selected $x_L$ values.  The separation between each pair of dashed horizontal lines in this image is $0.02$.
%
Following from Eq.\,\eqref{ASSA}, the magnitude of the SSA is roughly $\rho \bar\mu/(2+\lambda)$.
}
\end{figure}

\section{Predictions -- polarised target}
Figure~\ref{ResultsPol} illustrates our parameter-free predictions for the angular distribution coefficient, $\bar\mu$ in Eq.\,\eqref{DYcross}, associated with $\pi N$ Drell-Yan processes involving a longitudinally polarised target.
The other such coefficient, $\bar \nu$, is a simply proportional quantity -- see Eq.\,\eqref{barnu}.
The solid purple curve in each panel was obtained using the proton helicity-independent DFs calculated in Ref.\,\cite{Lu:2022cjx} and the helicity-dependent DFs calculated in Ref.\,\cite{Cheng:2023kmt}, evolved to $\zeta_5$ according to the all-orders DGLAP evolution scheme detailed in Ref.\,\cite{Yin:2023dbw}, and the pion DA computed in Ref.\,\cite{Chang:2013pq}, evolved $\zeta_2\to \zeta_5$ as described in Sec.\,\ref{SecUP}.

The marked effect of correcting Eq.\,\eqref{corrected} is apparent when comparing the images in Fig.\,\ref{ResultsPol} with those in Ref.\,\cite[Fig.\,6]{Bakulev:2007ej}: herein, the large-$x_L$ sensitivity to the DA is much reduced.
(The curves in Ref.\,\cite[Fig.\,6]{Bakulev:2007ej} also have the wrong sign.)

Regarding the panels in Fig.\,\ref{ResultsPol}, it will be seen that whilst $\bar\mu$ does show a modest quantitative sensitivity to $s$, its qualitative behaviour is stable.  Column~3 in Fig.\,\ref{ResultsPol} highlights that the behaviour of $\bar\mu$ is influenced by the $x$-dependence of the pion DA.
However, similar ratios for $\lambda$ cover the same range of magnitudes.
Hence, as with the unpolarised angular distribution coefficients, precise data would be needed in order to use $\bar\mu$ as a discriminator between the pointwise behaviour of distinct yet simultaneously realistic pion DAs.

\section{Angular moment of polarised target cross-section}
In Ref.\,\cite{Brandenburg:1995pk} it was suggested that the following angular average of the Drell-Yan cross-section:
\begin{align}
{\cal M}_{\rm ang} &= \int_0^\pi \! d\theta \int_0^{2\pi}\!\! d\phi \, \sin 2\theta\, \sin \phi\,
 \frac{d^5 \sigma(\pi^- N \to \mu^+ \mu^- X)}{dQ^2 dQ_{T}^2 dx_L d\cos\theta d\phi} \nonumber \\
& \propto  N(\tilde x, \rho)  \bar\mu \propto  - 2\pi \bar\mu_N \rho \tilde x F_\varphi \varphi(\tilde x, Q^2)\,,
\label{EqMang}
\end{align}
may be especially sensitive to the $x$-dependence of the pion DA because it is directly proportional to $\varphi_\pi(x)$.

Our predictions for ${\cal M}_{\rm ang}$, identified with the right-hand side of Eq.\,\eqref{EqMang}, are exemplified in Fig.\,\ref{ResultsMang}, wherein they are also compared with results obtained using the other DAs in Fig.\,\ref{DAcompare}.  These images indicate that, despite earlier suggestions \cite{Brandenburg:1995pk, Bakulev:2007ej}, ${\cal M}_{\rm ang}$ is not a better discriminator between realistic pion DAs than $\bar\mu$ itself: equally precise data would be needed in both cases.
On the other hand, combining comparisons against both $\bar\mu$ and ${\cal M}_{\rm ang}$ would be valuable because these quantities express distinct $x$-dependent signatures of differences between realistic pion DAs.

\section{Single-spin asymmetry}
The reaction mechanism drawn in Fig.\,\ref{DrellYan} leads to a nonzero single-spin azimuthal asymmetry (SSA):
\begin{align}
{\cal A} & = \frac{ d\sigma({\mathpzc s}_L = +1) - d\sigma({\mathpzc s}_L = -1)}
{d\sigma({\mathpzc s}_L = +1) + d\sigma({\mathpzc s}_L = -1)}\,.
\end{align}
Integrating over the polar angle, one obtains
\begin{align}
\label{ASSA}
{\cal A}(\phi,x_L,\rho) & = \frac{\rho \sin 2\phi \, \bar\mu({\mathpzc s}_L = +1)}{2 + \lambda + \tfrac{1}{2}\nu \cos 2\phi}\,.
\end{align}
Evidently, projected this way, the SSA is a sinusoidal function whose magnitude increases with $\rho$ and is modulated by $\bar\mu({\mathpzc s}_L = +1)$.
As remarked elsewhere \cite{Bakulev:2007ej}, ${\cal A} \neq 0$ follows from the fact that the nucleon's partons are able to transfer their polarisation to the azimuthal distribution of the $\mu^+ \mu^-$ pair because the reaction in Fig.\,\ref{DrellYan} features an imaginary part and possible phase interference between contributing amplitudes owing to the presence of the pion DA and associated hard-gluon exchange.  Absent these bound state effects, such SSAs require additional perturbative radiative corrections and $T$-odd parton DFs \cite{Sivers:1989cc, Brodsky:2002cx, Collins:2002kn}.

Our predictions for ${\cal A}(\phi,x_L,\rho)$ are illustrated in Fig.\,\ref{FigSSA}.  This quantity does not exhibit material sensitivity to the pion DA beyond the fact that its incorporation is essential for a nonzero value.  Comparison with Ref.\,\cite[Fig.\,10]{Bakulev:2007ej} reveals again the impact of correcting Eq.\,\eqref{corrected}.

\section{Conclusions}
Supposing Fig.\,\ref{DrellYan} to be a sound representation of the $\pi^- N \to \mu^+ \mu^- X$ process and using contemporary predictions for proton parton distributions, we have confirmed that, on the valence antiquark domain, the measured $\mu^+$ angular distributions should be mildly sensitive to the pointwise form of the pion distribution amplitude (DA), $\varphi_\pi(x)$.  Contrary to some earlier suggestions, however, polarised target experiments do not deliver a significant sensitivity improvement over those with unpolarised targets.  They may even be less favourable given the increased difficulty of such experiments and typically lower precision of the data collected.

When attempting to exploit this sensitivity in drawing a map of $\varphi_\pi(x)$, one finds that the only currently available $\pi N \to \mu^+ \mu^- X$ data were obtained using an unpolarised target; all are more than thirty years old; and none of the information inferred therefrom is of sufficient precision to enable its use as a discriminator between pion DAs that, by some measures, are empirically equivalent.  Nevertheless, the available information strongly favours a pion DA that is markedly dilated in comparison with the asymptotic form.  Such dilation can be seen as an expression of emergent hadron mass in the Standard Model \cite{Binosi:2022djx, Ding:2022ows, Ferreira:2023fva, Carman:2023zke}.  It may be anticipated that precise data from new generation experiments \cite{Anderle:2021wcy, Arrington:2021biu, Quintans:2022utc, Accardi:2023chb} will serve to harden this case.

A curious additional feature of the bound-state-sensitive reaction mechanism in Fig.\,\ref{DrellYan} is that it is sufficient to produce a nonzero single-spin azimuthal asymmetry (SSA), without reference to $T$-odd parton distribution functions (DFs).  This possibility may complicate the extraction of such $T$-odd DFs from data, requiring that additional attention be paid to identifying those kinematic domains upon which such DFs are the dominant source of a given SSA.

\medskip
\noindent\textbf{Acknowledgments}.
%
%
We are grateful for constructive comments from O.~Denisov, J.~Friedrich, W.-D.~Nowak,
S.~Platchkov, C.~Quintans, K.~Raya and J.~Rodr\'iguez-Quintero.
Work supported by:
National Natural Science Foundation of China (grant no.\,12135007);
Natural Science Foundation of Jiangsu Province (grant no.\ BK20220323);
and
Helmholtz-Zentrum Dresden Rossendorf under the High Potential Programme.


\medskip
\noindent\textbf{Declaration of Competing Interest}.
The authors declare that they have no known competing financial interests or personal relationships that could have appeared to influence the work reported in this paper.



\end{document}